\documentclass[sn-mathphys]{sn-jnl}

\jyear{2023}%
\usepackage{framed} 
\usepackage{multicol} 
\usepackage{nomencl} 
\usepackage{amsmath}
\usepackage{amssymb}
\usepackage{subfig}
\usepackage{amsmath}
\usepackage{mathrsfs}
\usepackage{graphicx}
\usepackage{algorithm}
\usepackage{algpseudocode}
\usepackage{float}
\usepackage{multirow}
\usepackage[utf8]{inputenc}
\usepackage{lineno}
\usepackage{arydshln}
\usepackage{adjustbox}

\theoremstyle{thmstyleone}%
\theoremstyle{thmstyletwo}%

\theoremstyle{thmstylethree}%

\begin{document}

\title[The potential impact of large-scale wind clusters on the local weather patterns]{The potential impact of large-scale wind clusters on the local weather patterns}


\author[1]{\fnm{Rui} \sur{Li}}\email{rui.li.8@warwick.ac.uk}

\author[1]{\fnm{Jincheng} \sur{Zhang}}\email{jincheng.zhang.1@warwick.ac.uk}

\author*[1]{\fnm{Xiaowei} \sur{Zhao}}\email{xiaowei.zhao@warwick.ac.uk}

\affil[1]{\orgdiv{Intelligent Control \& Smart Energy (ICSE) Research Group, School of Engineering}, \orgname{University of Warwick}, \orgaddress{\city{Coventry},\country{UK}}}




\abstract{To decarbonise the electricity sector and achieve renewable energy targets, a rapidly growing number of wind farms have been authorised, constructed, and commissioned in the UK and EU in recent years. For instance, the UK Government aims to expand offshore wind capacity to 60 GW by 2030, while the EU has set a target of 120 GW of offshore renewable energy by the same year. Given these substantial projected capacities, it is crucial to thoroughly investigate the potential impacts of large-scale wind clusters on local weather patterns to prevent unintended consequences prior to deployment. In this paper, we use the WRF model to simulate four scenarios with varying wind energy capacities in the North Sea, assessing the potential effects of these wind clusters on the local weather patterns over mainland UK. Please note that the simulations of Case 3 and Case 4 are still ongoing, while all analyses in the current version of manuscript are all based on Case 1 and Case 2.}




\maketitle
 
\section{Main}\label{sec1}
Offshore wind, as a promising carbon-free and sustainable source of energy, has seen rapid and intensified exploitation in recent years, fuelled by the sharp rise in global energy prices and the urgent need to meet net-zero targets to combat climate change \cite{perera2023challenges, howland2022collective, cherp2021national}. For example, the UK government is committed to quadruple offshore wind by 2030 to deliver up to 60 GW capacity. Meanwhile, to achieve the UK's net zero targets by 2050, the Sixth Carbon Budget \cite{climate2020sixth} published by the Climate Change Committee recommends the deployment of renewables at scale, sustaining that build rate to support the deployment of up to 125 GW of offshore wind by 2050. Similarly, the Ostend Declaration of Energy Ministers outlines the EU’s plan to deploy around 120 GW of offshore wind by 2030 and at least 300 GW by 2050 in the North Sea. With thousands of turbines projected and hundreds of farms planned, the potential impact of such large-scale wind clusters on the local weather pattern stands out as an important yet inconclusive issue which needs comprehensive and in-depth investigation before the actual deployment of the projects. 

\begin{figure}[htb]
\includegraphics[width=11.8cm]{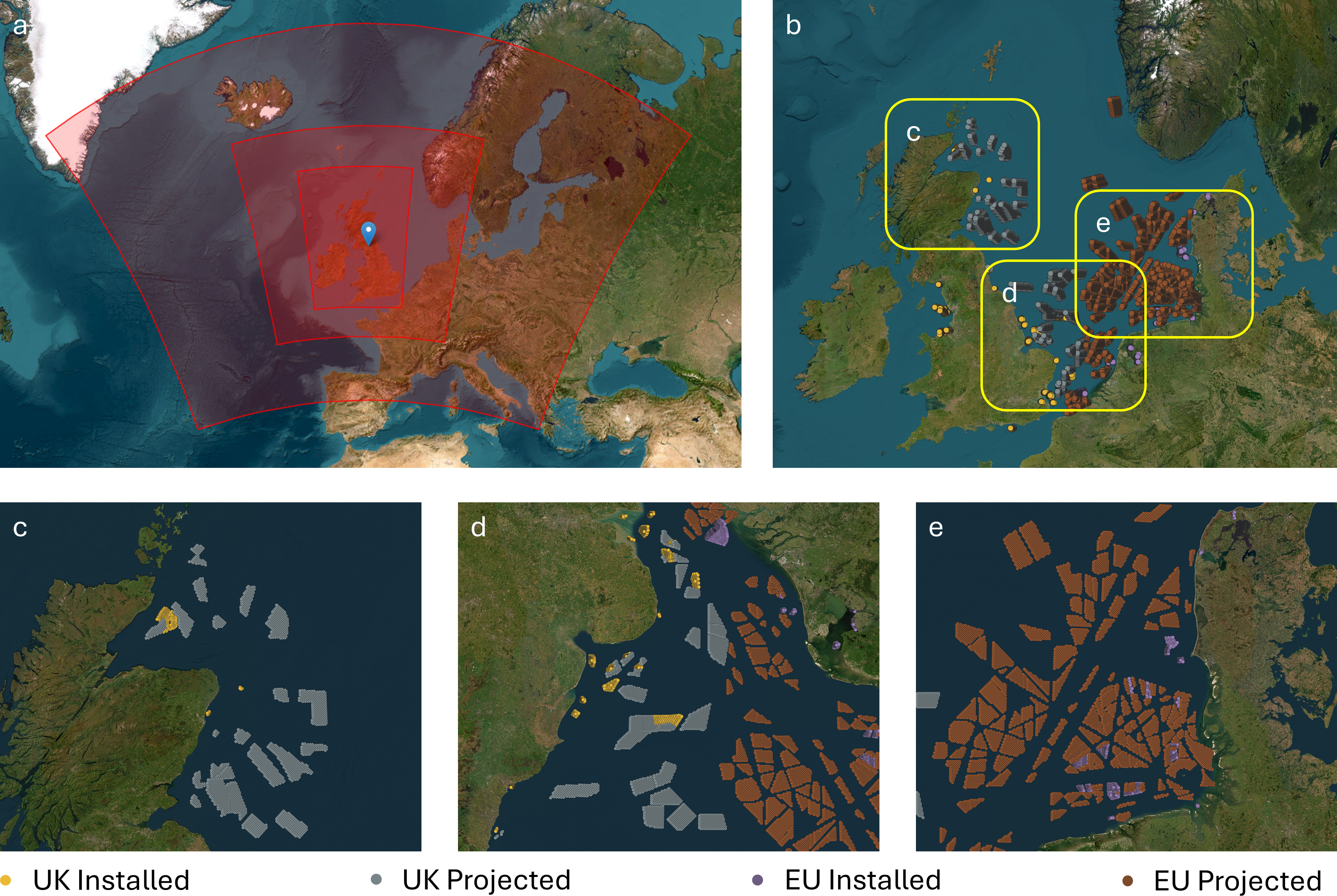}
\caption{The (a) WRF simulation domain and (b) included wind turbines in our study, where the corresponding turbine-intensive areas in (b) are enlarged in (c)-(e).}
\label{fig:1}
\end{figure}

Current understanding of the environmental impacts of wind farms primarily stems from studies focused on onshore installations \cite{ouro2024environmental, zhou2020weather, wu2021near}. These studies reveal a range of effects, such as on local land surface temperatures. For instance, several studies have shown that wind farm installations can lead to a rise in local temperatures, especially during the night \cite{liu2022heterogeneous, zhang2023impact, miller2018climatic, xia2016case, zhou2013diurnal}, with satellite observations confirming this phenomenon \cite{harris2014satellite}. However, these temperature impacts tend to vary based on land use types, with different areas experiencing contrasting effects \cite{liu2023remotely}. In terms of ecological impacts, some research has indicated that wind farms can negatively affect plant biomass production, reducing vegetation growth in affected areas \cite{tang2017observed, gao2023impact}. Interestingly, other studies have suggested that large-scale renewable installations, such as wind and solar farms, can enhance local rainfall, as demonstrated in simulations over the Sahara \cite{li2018climate}. This increased rainfall has the potential to boost vegetation cover \cite{li2018climate}. Similarly, mesoscale simulations in the USA have reported increased precipitation due to wind farms \cite{fiedler2011effect}. On the agricultural front, wind farms may have positive spillover effects, with evidence suggesting increased crop yields, such as corn, in areas near wind farm developments \cite{kaffine2019microclimate}. Despite these findings, the regional and broader-scale impacts of wind farms remain a topic of debate. Some studies, using the Weather Research and Forecasting (WRF) Model, have found that while the impacts of wind farms are significant and sustained at a local level, they tend to be minor and occasional on a regional scale \cite{wang2019impact, pryor2018influence}. High-resolution WRF simulations over the eastern USA further support this, showing that the climate impacts of current and future wind turbine deployments, necessary for generating 20\% of the region's electricity, are modest when compared to the impacts of land use changes or coal-based power generation for the same energy output \cite{pryor202020}. Similarly, simulations with the Community Atmosphere Model (CAM) have shown that the climate impacts of wind energy, even at capacities as high as 20 TW, are largely negligible beyond the immediate areas surrounding the wind farms \cite{fitch2015climate}. However, other research presents a contrasting view, highlighting more pronounced regional effects. For example, large-scale wind farm installations have been linked to significant reductions in wind speed, regional temperature changes, and upper-atmosphere variations, as observed in mesoscale numerical simulations \cite{wang2023climatic, sun2018impacts, huang2019numerical}. Additionally, wind farms have been found to cause notable reductions in soil moisture within and around the farms, which could exacerbate drought conditions and negatively affect grassland ecosystems \cite{wang2023wind}. The impacts caused by the construction procedure within the forested areas have also been investigated, such as the reduced vegetation cover and increased soil erosion  \cite{xia2025assessment}.

When it comes to offshore wind farms, the existing body of literature remains relatively limited \cite{ouro2024environmental}. Offshore wind turbines directly impact the atmosphere by creating additional drag and generating wake turbulence \cite{vautard2014regional}. This turbulence enhances surface-level mixing, leading to a reduced daily temperature range, with cooler daytime temperatures and warmer nighttime temperatures \cite{baidya2010impacts}. In areas with dense wind farm coverage, remote sensing data has revealed a net warming effect of approximately 0.7°C over a decade \cite{zhou2012impacts, zhou2013diurnal, rajewski2013crop}, findings that are corroborated by mesoscale model simulations \cite{roy2011simulating, baidya2004can}. Wind tunnel experiments also suggest that wind farms alter the energy budget in their wake \cite{zhang2013experimental}, and the turbulence generated by wind turbines influences the entire planetary boundary layer, causing localised modifications to airflow patterns around the farms \cite{rajewski2013crop, fitch2012local, smith2013situ}. The wind power resource itself can be affected when the extracted power density exceeds 1 W/m², potentially reducing wind speeds \cite{miller2011estimating, adams2013global}. Simulations using the Community Climate Model (CCM) Version 3 show that large-scale wind energy installations may cause surface warming exceeding 1°C over land, while surface cooling of over 1°C is observed over ocean installations when 10\% or more of global energy is generated by wind by 2100 \cite{wang2010potential}. However, further simulations suggest that offshore deployments cause only minor perturbations to the global large-scale atmospheric circulation \cite{wang2011potential}. WRF simulations conducted in the USA show a reduction in hub-height wind speeds within and downwind of wind farms due to wake effects \cite{golbazi2022surface, quint2024meteorological}. However, meteorological changes, such as variations in 2-metre temperatures and surface heat fluxes induced by offshore wind turbines, are nearly imperceptible \cite{golbazi2022surface}. Airborne meteorological in-situ measurements from the German Bight reveal a significant increase in vertical latent heat fluxes, especially in the upward direction, over offshore wind farm clusters \cite{platis2023impact}. In addition to the observed wind speed reductions, a one-year evaluation using the HARMONIE-AROME numerical weather prediction model shows that temperature and humidity profiles are altered due to enhanced turbulent mixing by the turbines, particularly under stable atmospheric conditions \cite{van2022one}. Offshore wind farms can also affect extreme weather events, as demonstrated by studies on tropical cyclones, which show changes in precipitation, wind speeds, and low-level moisture convergence due to large-scale offshore installations \cite{zhang2024impacts}. Coastal impacts have also been observed, such as reduced upwelling on the coastal side of wind farms and increased upwelling on the offshore side, as seen along the California coast \cite{raghukumar2023projected}. The ecological impacts of offshore wind farms have been explored as well. For instance, studies have examined the effects on scallop larvae dispersal and settlement in the US Northeast Shelf region \cite{chen2024potential}, as well as the spatial and temporal impacts on fish movement and habitat changes in Europe \cite{van2020effects}.

In summary, the potential impacts of offshore wind farm deployment remain highly uncertain and ambiguous due to the current lack of large-scale, long-term, and high-resolution simulations. Committing to and deploying large-scale wind clusters with significant uncertainty and unpredictability could engender substantial risks to the region's climate and ecology. To address such huge uncertainty and ambiguity, we conduct a three-year long high-resolution WRF simulation covers the UK mainland and adjacent waters such as the North Sea, the Irish Sea and the Celtic Sea where a series of extensive wind farms have been installed and planned. Four case studies with projected 25.741 GW, 103.843 GW, 210.849 GW and 381.701 GW offshore wind capacity within the simulation domain are implemented, compared and analysed, with corresponding UK capacities of 13.270 GW, 51.124 GW, 101.248 GW, and 146.788 GW. The simulation domain and the included wind turbines are illustrated in Fig. \ref{fig:1}.

\section{Experimental setting and results}\label{sec2}

\begin{figure}[htb]
\centering
\includegraphics[width=10cm]{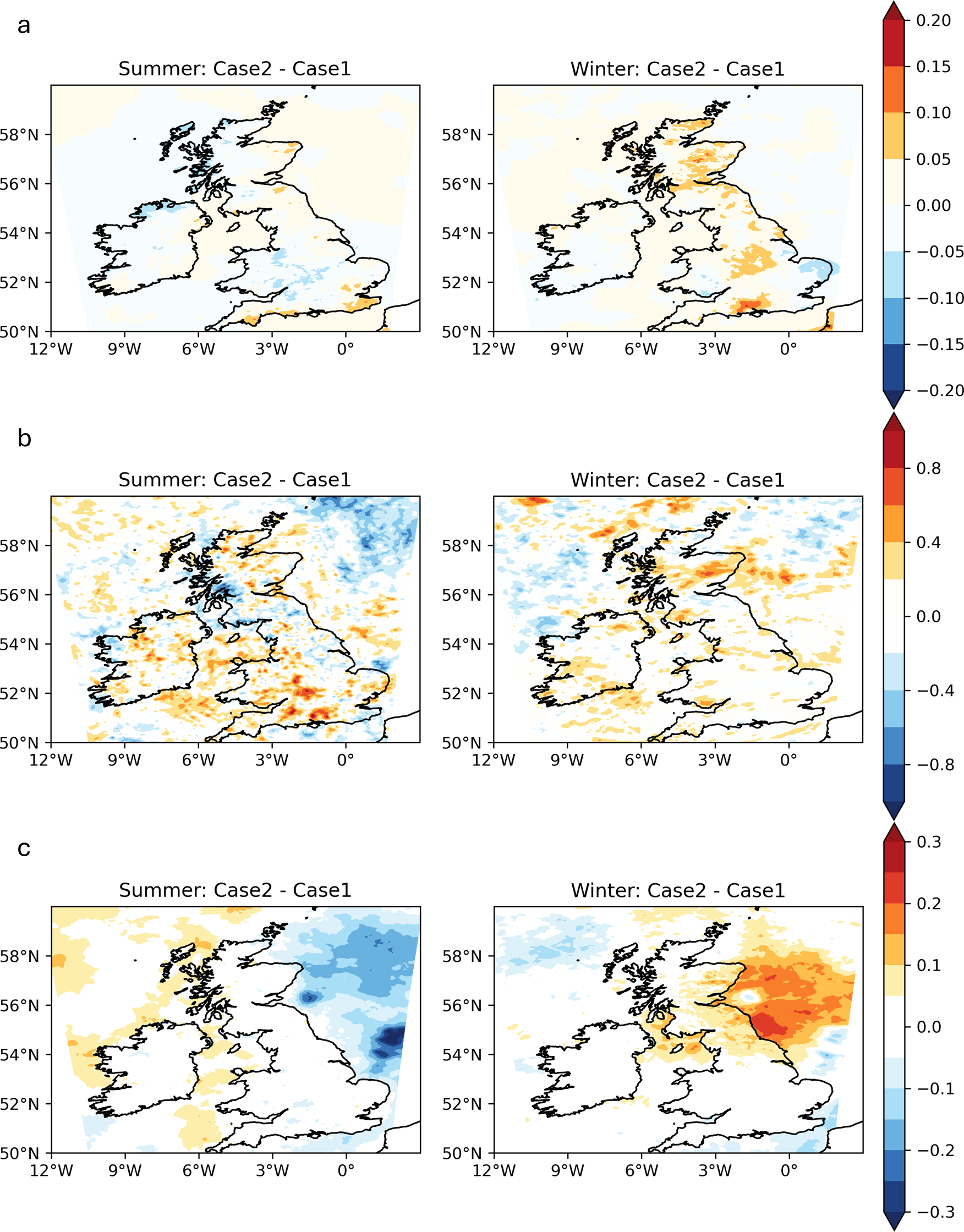}
\caption{The average difference of (a) temperature, (b) precipitation and (c) wind speed within Domain 3 for summer and winter between Case 1 and Case 2.}
\label{fig:2}
\end{figure}

\begin{table}[htb]
\setlength{\abovecaptionskip}{0.cm}
\caption{The number and power of simulated turbines under four cases.}
\label{table:simulation_case}
\begin{adjustbox}{center}
\begin{tabular}{ccccccc}
\hline
\multicolumn{3}{l}{} & case1 & case2 & case3 & case4 \\ \hline
\multirow{6}{*}{UK} & \multirow{2}{*}{Baseline} & Turbines & 2860 & 2860 & 2860 & 2860 \\
& & Power (MW) & 13270 & 13270 & 13270 & 13270 \\\cdashline{2-2}[1pt/1pt]
& \multirow{2}{*}{18 MW} & Turbines & 0 & 2103 & 0 & 0 \\
& & Power (MW) & 0 & 37854 & 0 & 0 \\\cdashline{2-2}[1pt/1pt]
& \multirow{2}{*}{22 MW} & Turbines & 0 & 0 & 3999 & 6069 \\
& & Power (MW) & 0 & 0 & 87978 & 133518 \\\cdashline{2-2}[1pt/1pt]
& \multirow{2}{*}{Total} & Turbines & 2860 & 4963 & 6859 & 8929 \\
& & Power (MW) & 13270 & 51124 & 101248 & 146788 \\ \hdashline
\multirow{8}{*}{EU} & \multirow{2}{*}{Baseline} & Turbines & 2453 & 2453 & 2453 & 2453 \\
& & Power (MW) & 12471 & 12471 & 12471 & 12471 \\\cdashline{2-2}[1pt/1pt]
& \multirow{2}{*}{18 MW} & Turbines & 0 & 2236 & 0 & 0 \\
& & Power (MW) & 0 & 40248 & 0 & 0 \\\cdashline{2-2}[1pt/1pt]
& \multirow{2}{*}{22 MW} & Turbines & 0 & 0 & 4415 & 10111 \\
& & Power (MW) & 0 & 0 & 97130 & 222442 \\\cdashline{2-2}[1pt/1pt]
& \multirow{2}{*}{Total} & Turbines & 2453 & 4689 & 6868 & 12564 \\
& & Power (MW) & 12471 & 52719 & 109601 & 234913 \\ \hdashline
\multirow{2}{*}{UK + EU} & \multirow{2}{*}{Total} & Turbines & \textbf{5313} & \textbf{9652} & \textbf{13727} & \textbf{21493} \\
& & Power (MW) & \textbf{25741} & \textbf{103843} & \textbf{210849} & \textbf{381701} \\ \hline
\end{tabular}
\end{adjustbox}
\end{table}

In our simulation, four scenarios with varying offshore wind capacity deployments are modelled, with the WRF model configuration detailed in Section \ref{secA}. Case 1, serving as the baseline, represents the offshore turbine deployment status as of 2021, with approximately 13 GW in the UK and 12 GW in the EU within the simulation domain. The remaining three cases simulate future deployment scenarios with offshore wind capacities of approximately 104 GW, 211 GW, and 382 GW, respectively. The number of turbines and total power capacities for each scenario are detailed in Table \ref{table:simulation_case}. The simulations were conducted on the AWS platform using a 192-core hpc7a.96xlarge instance, with each case requiring approximately 1,200 CPU hours to complete.

\begin{figure}[htb]
\includegraphics[width=12cm]{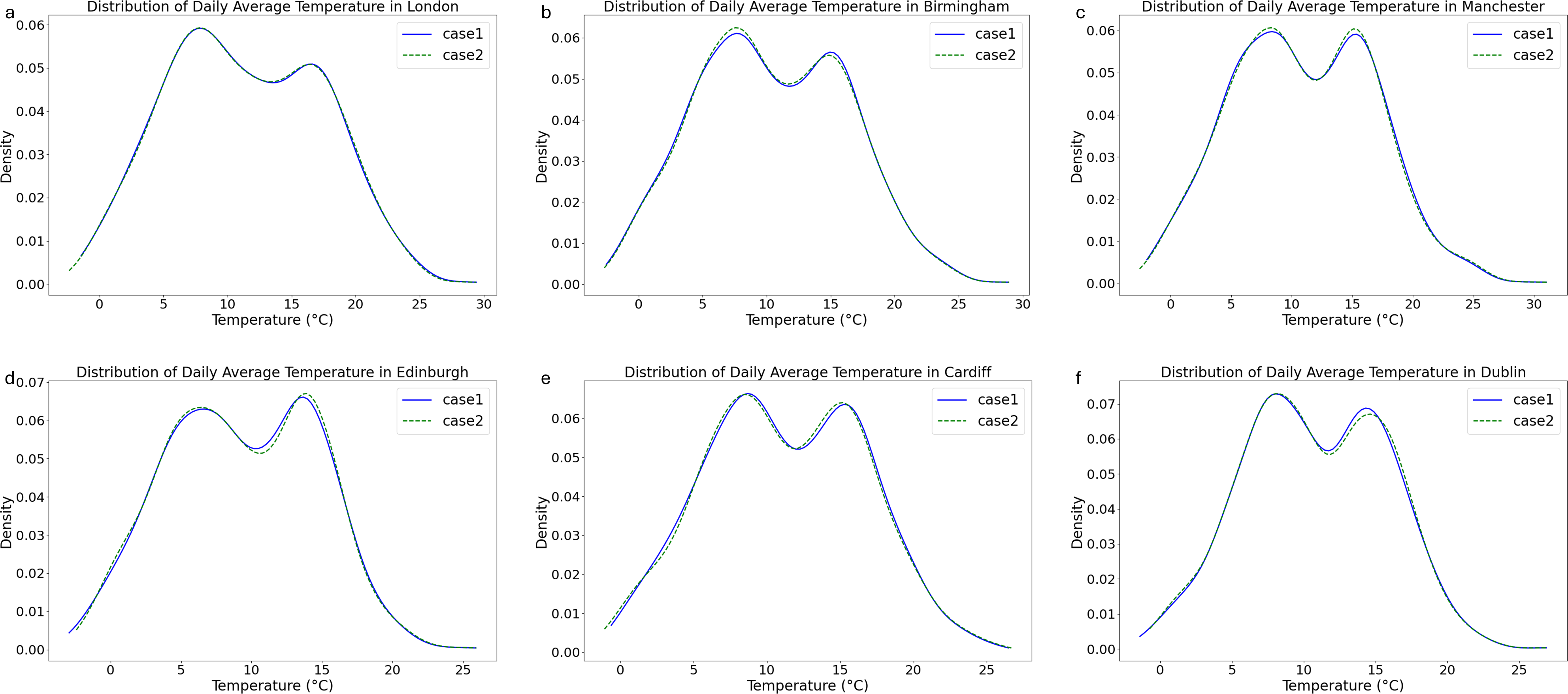}
\caption{The distribution of daily average temperature along three years under case 1 and case 2 over (a) London, (b) Birmingham, (c) Manchester, (d) Edinburgh, (e) Cardiff and (f) Dublin.}
\label{fig:3}
\end{figure}

\subsection{Atmospheric variations}\label{sec2.1}
To visualise and compare the atmospheric variations across different deployment scenarios, the differences in temperature, precipitation, and wind speed are presented in Fig. \ref{fig:2}, while the differences in average daily minimum and maximum temperatures are shown in Fig. \ref{fig:C1}. For temperature, the average summer differences are minimal overall. However, most of the eastern coastal regions of the UK experience a noticeable increase in daily maximum temperature as shown in Fig. \ref{fig:C1} (b). In contrast, a decrease in the average summer daily maximum temperature is observed in Gloucestershire, Oxfordshire, and Warwickshire in England, as well as in the Highlands of Scotland, also shown in Fig. \ref{fig:C1} (b). During winter in Fig. \ref{fig:2} (a), slight increases are observed in the heartland of England and approximately half of Scotland, with the most significant change occurring in Hampshire, where temperatures increase by over $0.2,^\circ\mathrm{C}$. For precipitation in Fig. \ref{fig:2} (b), the North Sea shows reduced rainfall, particularly in summer. On the UK mainland, increased precipitation is observed in Gloucestershire, Hampshire, and Dorset during summer, and in Aberdeenshire during winter. For wind speed in Fig. \ref{fig:2} (c), the differences exhibit stark seasonal contrasts. In summer, the North Sea experiences a clear reduction in average wind speeds, while in winter, an evident increase is observed.

\begin{figure}[htb]
\includegraphics[width=12cm]{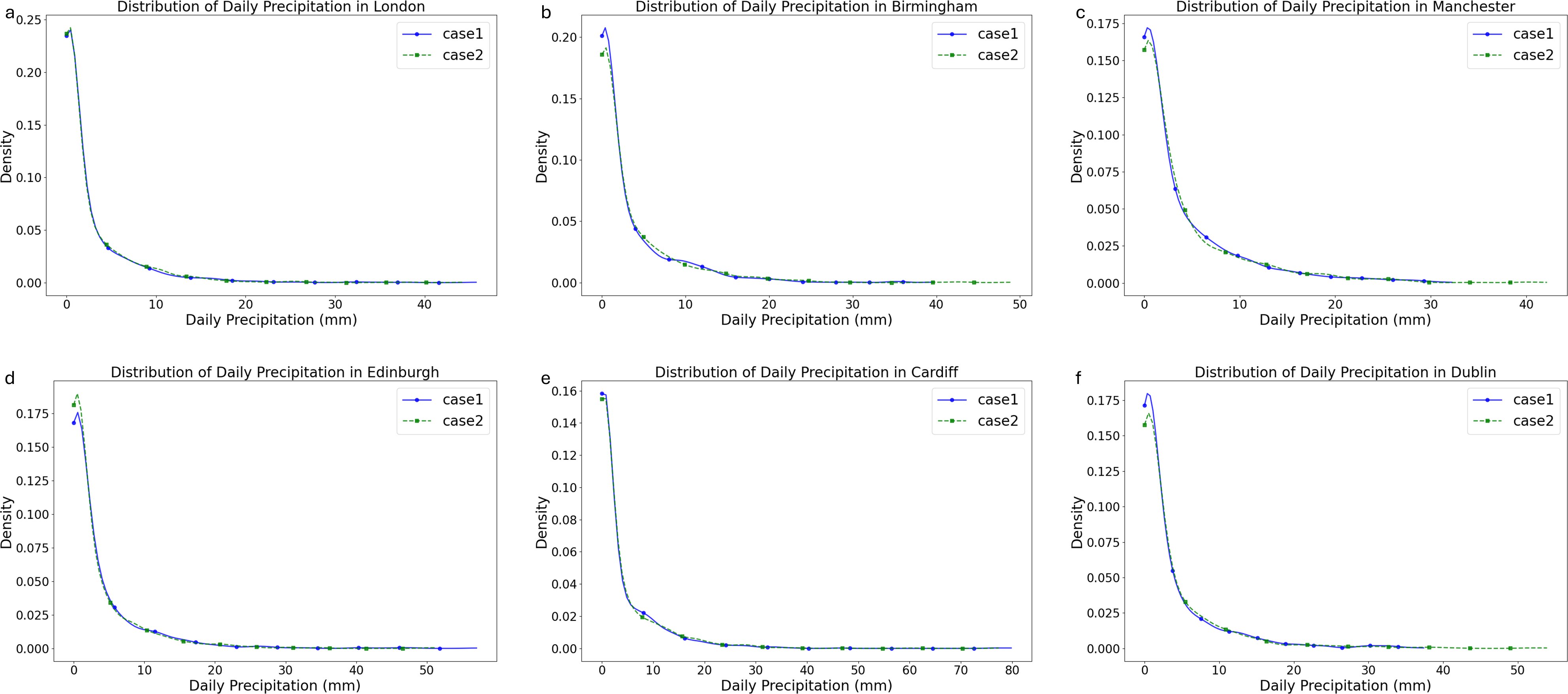}
\caption{The distribution of daily average precipitation along three years under case 1 and case 2 over (a) London, (b) Birmingham, (c) Manchester, (d) Edinburgh, (e) Cardiff and (f) Dublin.}
\label{fig:4}
\end{figure}

\subsection{Impact on the large cities}\label{sec2.2}

\begin{figure}[htb]
\includegraphics[width=12cm]{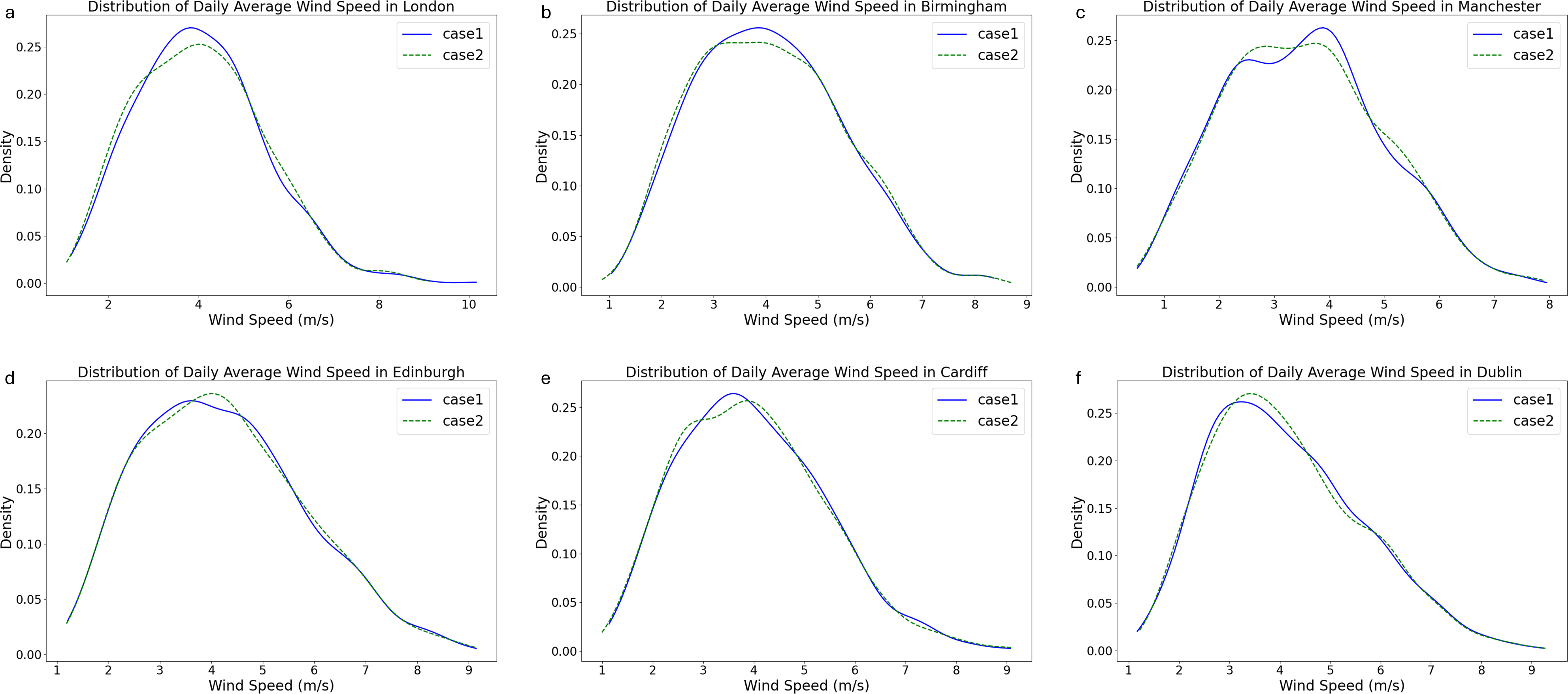}
\caption{The distribution of daily wind speed along three years under case 1 and case 2 over (a) London, (b) Birmingham, (c) Manchester, (d) Edinburgh, (e) Cardiff and (f) Dublin.}
\label{fig:5}
\end{figure}

In this section, we compare the three-year averages of temperature, precipitation, and wind speed for London, Birmingham, Manchester, Edinburgh, Cardiff and Dublin included in Domain 3 under two different deployment scenarios. The results are illustrated in Fig. \ref{fig:3}, Fig. \ref{fig:4}, and Fig. \ref{fig:5}. For the daily average 2 m temperature, all cities show a mild and slight impact from the deployment of large-scale offshore wind clusters. For example, London experiences slightly more colder days, while Edinburgh becomes marginally warmer. In the case of daily precipitation, the impacts are more pronounced. Birmingham, Manchester, and Dublin, in particular, see an increase in extreme conditions, with a higher frequency of both very low and very high precipitation events. These findings suggest the need for better preparedness to manage such weather extremes. For the daily average 10 m wind speed, the effects are concentrated in the mild wind range of 3 m/s to 6 m/s, with some noticeable variations observed. Detailed variations in daily averages of wind speed, temperature, and precipitation for these six cities are provided in Section \ref{secC}. Notably, under the 50 GW deployment scenario, Birmingham and Manchester experience more frequent extreme precipitation events, with daily totals exceeding 30 mm and occasionally surpassing 50 mm.

\subsection{North Sea wind speed changes}\label{sec2.3}
 
\begin{figure}[htb]
\includegraphics[width=12cm]{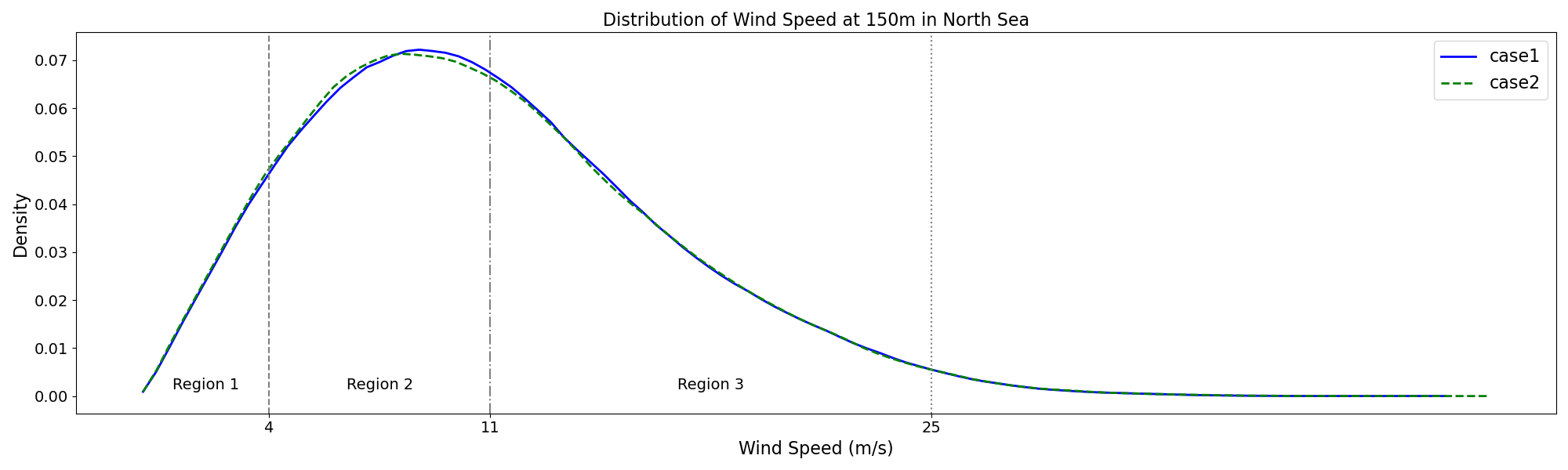}
\caption{The distribution of 150 m daily wind speed along three years under case 1 and case 2 over the North Sea.}
\label{fig:6}
\end{figure}

A significant concern regarding the deployment of large-scale offshore wind clusters is the potential reduction in hub-height wind speeds, which could decrease the available wind energy reserves. In this section, we visualise and compare the distribution of three-year wind speeds across the entire North Sea at 150 m height, as shown in Fig. \ref{fig:6}. Using the NREL 18 MW reference offshore wind turbine as an example, we mark the cut-in, rated, and cut-out wind speeds on the figure, as power generation is most sensitive within Region 2. The figure indicates that wind speeds within Region 2 experience a slight shift, with an increase in lower speeds and a reduction in higher speeds. This change, calculated across the entire North Sea, signifies a tangible impact on wind energy reserves, particularly within the critical Region 2 range. The influence appears to intensify under the higher-capacity scenarios of Case 3 and Case 4, which will be further analysed once the simulations are complete.

\section{Discussion}\label{sec5}
In this study, we investigate the potential impacts of large-scale offshore wind farm deployment on local climate patterns using WRF simulations. Four deployment scenarios are considered, though simulations for Case 3 and Case 4 are still ongoing. Preliminary comparisons between the 13 GW and 50 GW scenarios reveal measurable and recognisable impacts on temperature, precipitation, and wind speed. Specifically, slight shifts in climate patterns are observed over the UK mainland, alongside an increase in severe weather events in certain cities. Additionally, greater offshore wind farm deployment results in reduced wind resource availability across the North Sea, an important factor to consider for future planning and commissioning.

\begin{appendices}
\section{WRF Model Configuration}\label{secA}
\setcounter{figure}{0}

We ran the WRF model with a three-domain, two-way nested configuration consisting of a parent domain (Domain 1) with a horizontal grid resolution of 27 km, covering \(160 \times 120\) grid points. This parent domain contains Domain 2 as a nested subdomain, which has a resolution that is three times finer (9 km) and spans \(202 \times 205\) grid points. In turn, Domain 3 is nested within Domain 2 and has a further refined grid resolution of 3 km, covering \(295 \times 412\) grid points (Fig. \ref{fig:1}). In this two-way nesting setup, the model first solves equations for the grid cells in Domain 1, which provides boundary conditions for the finer grids in Domains 2 and 3. The model then calculates values for all grid points within Domains 2 and 3, with the fine-grid values subsequently replacing those in the coarser domains to ensure feedback of high-resolution information into the parent domain. All domains are configured with 62 vertical levels. The Fitch turbine parametrisation is enabled for Domain 2 and 3 as they cover the North Sea and the UK mainland, respectively. 

In our case, the initial and boundary conditions for Domain 1 are provided by the ERA5 reanalysis dataset, which offers comprehensive atmospheric data at a horizontal resolution of approximately 31 km \cite{copernicus2023era5}. The simulation period spans from 00:00 UTC on 31 December 2020 to 00:00 UTC on 31 December 2023, with input data intervals set at six hours to align with ERA5’s temporal granularity. To manage the accumulation of numerical errors and ensure model stability, the simulation is re-initialized every 7 days. Each re-initialization starts with a 24-hour spin-up period, allowing the model to establish a dynamic balance before transitioning to data analysis. This approach mitigates the effects of potential drift in the model, particularly for extended simulations, while maintaining high fidelity to the atmospheric conditions provided by ERA5. In addition, we applied four-dimensional grid nudging in the outermost domain to dampen higher-frequency fluctuations that often lead to instabilities. No nudging was applied to the Domain 1 and 2, where the wind farm parametrisation is active. The wakes generated by the wind farms in these fine-resolution domains can travel significant distances and may extend into the coarser outer domain. To minimize interference of the nudging with the wind turbine wakes in the coarse domain, we restricted nudging in the outer domain to levels above 600 m AMSL, corresponding to levels above the 31st model level.

\begin{table}[h!]
\centering
\caption{Details of the WRF model setup.}
\begin{tabular}{ll}
\toprule
\textbf{Parameter}                   & \textbf{Selection/Value}                 \\ \midrule
WRF model version                    & 4.3.3                                      \\
Simulation period                    & 31 Dec 2020 – 31 Dec 2023                 \\
Initial/boundary conditions          & ERA5 reanalysis, 6 h, 31 km resolution   \\
Re-initialization interval           & 7 days with 24-hour spin-up              \\
Wind farm parametrisation           & Fitch with TKE advection                 \\
Domain configuration                 & 3 domains with two-way nesting           \\
Domain 1 resolution                  & 27 km, 160 × 120 grid points             \\
Domain 2 resolution                  & 9 km, 202 × 205 grid points              \\
Domain 3 resolution                  & 3 km, 295 × 412 grid points              \\   
Land surface model (LSM)             & Noah-modified 21-category IGBP-MODIS     \\
Planetary boundary layer (PBL) scheme & MYNN2                                    \\
Shortwave radiation                  & RRTMG shortwave                          \\
Longwave radiation                   & RRTMG scheme                             \\
Sea surface temperature (SST) update & GHRSST, 0.01$^\circ$ resolution               \\     
Output frequency                     & 360 min, 180 min, 60 min \\ \bottomrule
\end{tabular}
\end{table}

\subsection{Land use and sea surface temperature modifications}\label{secA1}

\begin{figure}[htb]
\centering
\includegraphics[width=10cm]{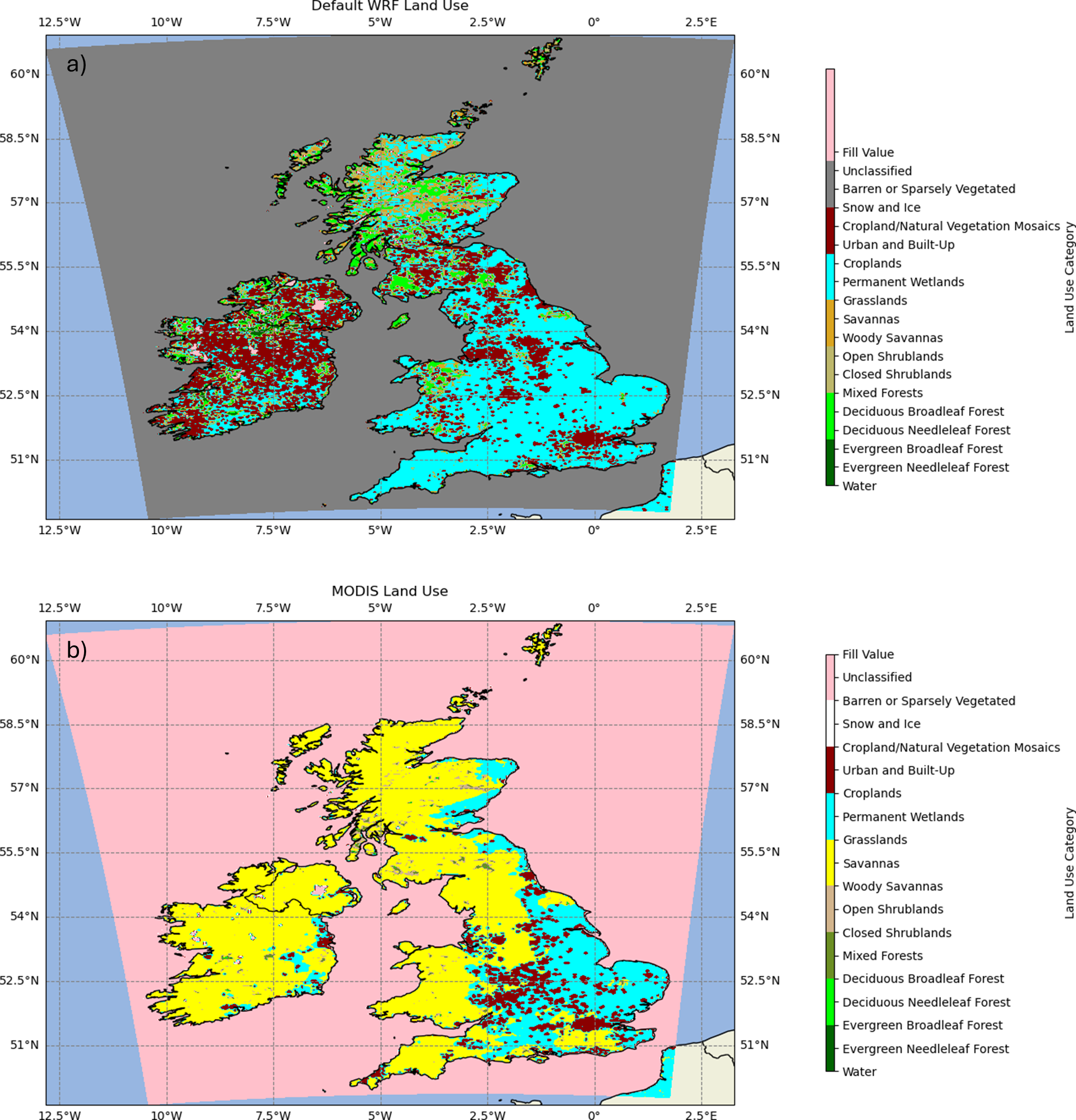}
\caption{The comparison between the MODIS and WRF default land use categories for Domain 1.}
\label{fig:landuse}
\end{figure}

The shape and land use of coastlines influence meteorology through land-atmosphere interactions at micro- and mesoscale levels. Accurate representation of coastlines is therefore essential to predict turbulent heat fluxes and other meteorological properties, including the timing and track of sea breeze circulations. To achieve this, we employed the 21-category IGBP-MODIS land use dataset \cite{friedl2022modis} to replace the default land use in the WRF model, which provides a detailed and up-to-date land classification suitable for regional modelling. The comparison between the MODIS and WRF default land use categories is illustrated in Fig. \ref{fig:landuse}.

In our setup, the sea surface temperature (SST) data is sourced from the GHRSST dataset \cite{reynolds2007daily} with a 0.01$^\circ$ ($\approx$  1 km) horizontal resolution. This high-resolution SST data improves upon the coarser SST provided by ERA5, allowing the model to resolve differential heating more accurately along coastlines and within coastal bays. The finer SST resolution captures temperature variations between coastal waters and open ocean, which is crucial for simulating local sea/land breeze characteristics. By using this detailed SST data, along with high-resolution MODIS land use information, the WRF model is better equipped to represent the land-sea contrasts and coastal dynamics, leading to improved accuracy in simulating coastal meteorological phenomena.

\subsection{WRF simulation verification}\label{secA2}

\begin{figure}[htb]
\centering
\includegraphics[width=10cm]{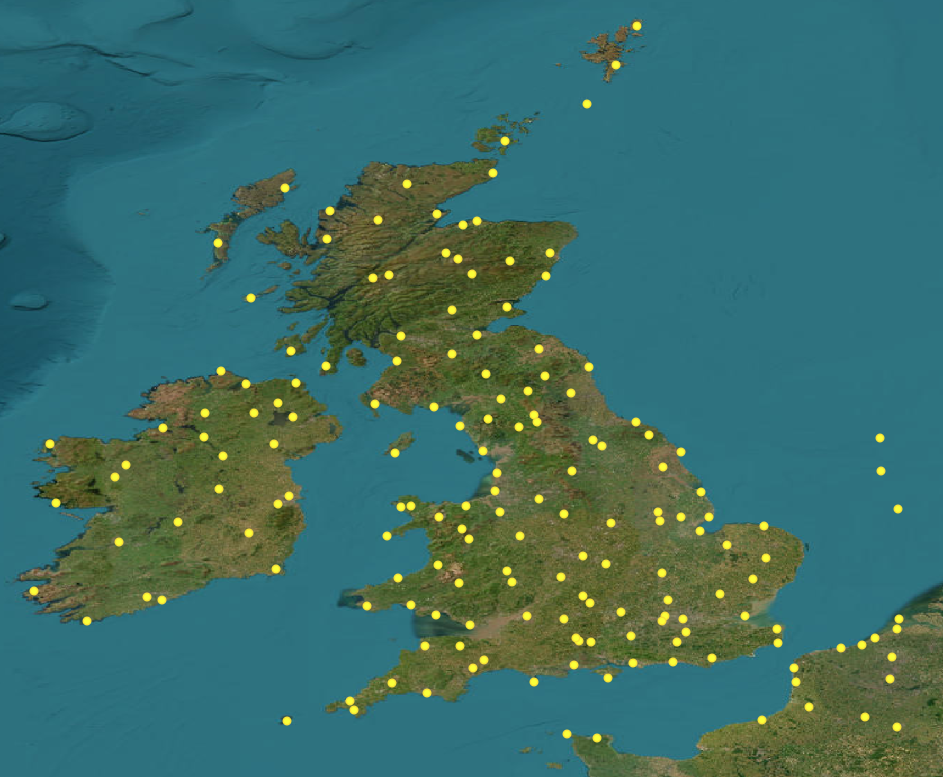}
\caption{The weather stations used to verify the accuracy of the WRF simulation.}
\label{fig:station}
\end{figure}

\begin{figure}[htb]
\centering
\includegraphics[width=12cm]{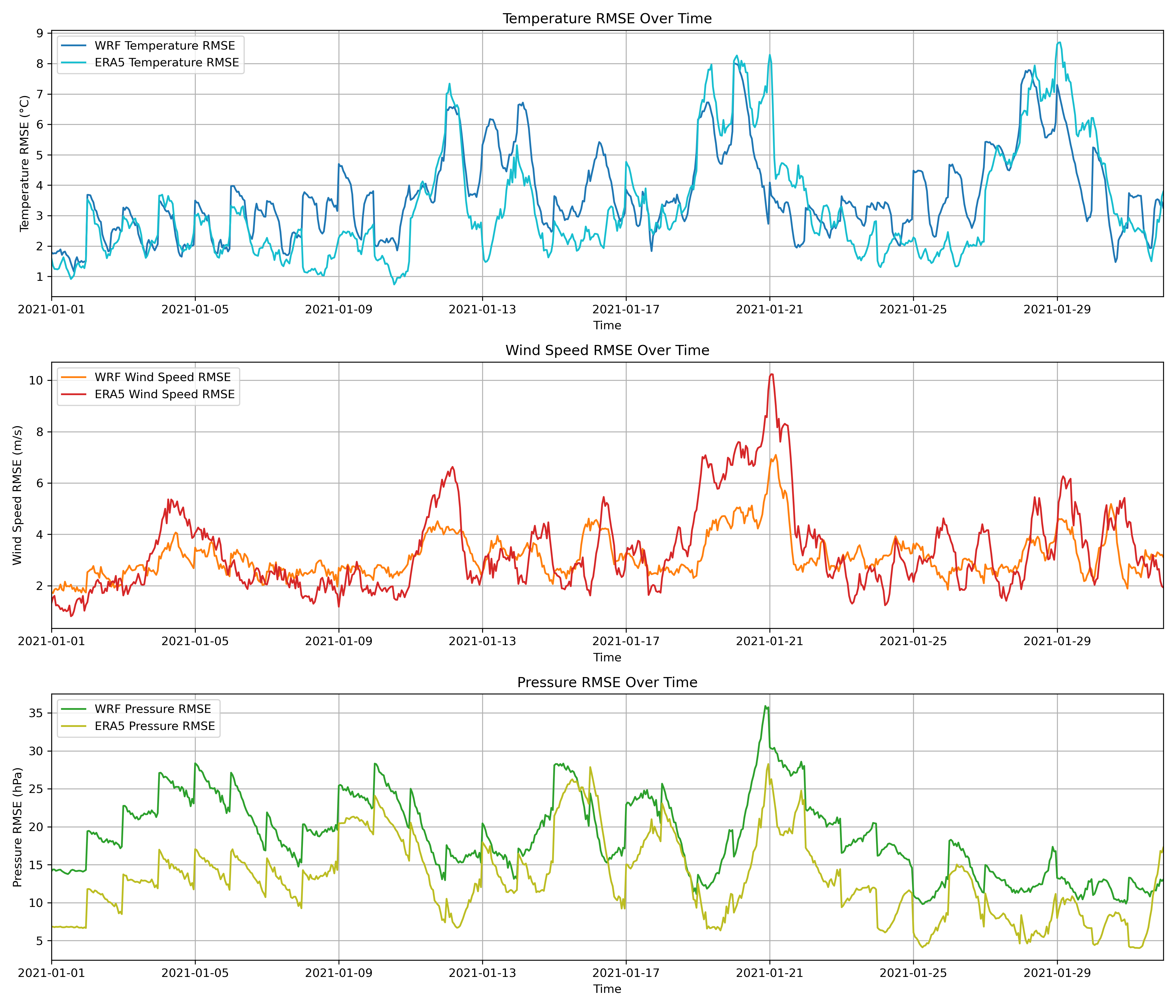}
\caption{The RMSE of the WRF simulation (case1) and ERA5 reanalysis data compared to the weather stations measurements in terms of temperature, 10 m wind speed and sea level pressure.}
\label{fig:rmse}
\end{figure}

To validate the accuracy of the simulation, measurement data from 169 weather stations within Domain 3 are collected, and the root mean square deviations (RMSE) of the WRF simulation (Case 1) and ERA5 reanalysis data are compared for January 2021. The locations of the stations are shown in Fig. \ref{fig:station}, and the RMSE comparison is illustrated in Fig. \ref{fig:rmse}.

The comparison demonstrates that the WRF simulation performs competitively with the ERA5 reanalysis data. Notably, for 10 m wind speed, the WRF simulation consistently shows better alignment with station measurements than the ERA5 data. These results confirm the accuracy and reliability of the WRF simulation for this study.

\section{Turbine parametrisation}\label{secB}
\setcounter{figure}{0}

In the WRF simulation, two types of offshore wind turbines are included: those already deployed and those projected for future deployment. To incorporate offshore turbines into the simulation, both precise coordinates and turbine power curves are required. For the existing turbines, locations are sourced from the 2021 global offshore wind turbine dataset \cite{zhang2021global}. However, due to proprietary restrictions, specific power curves for individual turbines are often unavailable. To address this, we compiled power curves for a range of reference turbines with capacities of 2 MW, 2.3 MW, 3.6 MW, 4 MW, 5 MW, 6 MW, 6.15 MW, 7 MW, 8 MW, 9.5 MW, 10.6 MW, 11 MW, 12 MW, 15 MW, 18 MW, and 22 MW \cite{zahle2024definition}. To assign a turbine with an approximate capacity, we calculated the average capacity by dividing the total power of a wind farm by the number of turbines within it, then selected the nearest matching power curve from our collection for all turbines in that farm. For projected turbines, we assume that advancements in turbine technology will lead to higher capacity installations. Thus, only the 18 MW and 22 MW power curves are used for future turbine parametrisation. Coordinates for these future installations are derived from plausible 2050 offshore wind locations in the North Sea \cite{waldman_plausible_2050_offshore_wind_2022}.

Case 1, representing the deployment status as of 2021, is used as a baseline, with the installed turbines included consistently across all cases for comparison. The three additional cases include varying numbers of 18 MW and 22 MW turbines to achieve the specified total offshore capacity targets for the UK—around 50 GW, 100 GW, and 150 GW, respectively, though the exact numbers may vary slightly. Detailed configurations for each case are provided in Table \ref{table:simulation_case}.

\section{Daily average variations}\label{secC}
\setcounter{figure}{0}

\begin{figure}[htb]
\centering
\includegraphics[width=10cm]{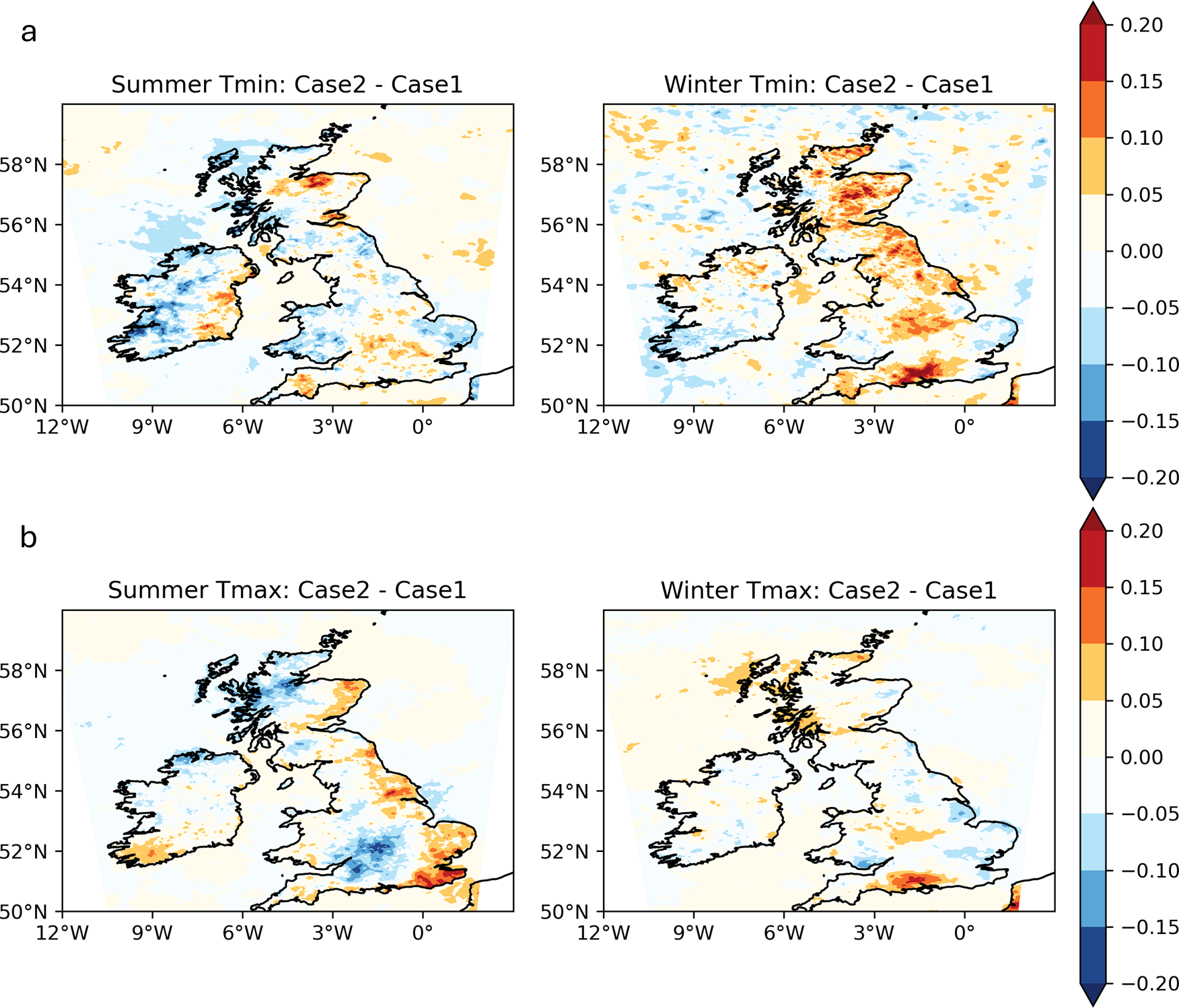}
\caption{The average difference of (a) daily minimum temperature and (b) daily maximum temperature within Domain 3 for summer and winter between Case 1 and Case
2.}
\label{fig:C1}
\end{figure}

\begin{figure}[htb]
\includegraphics[width=12cm]{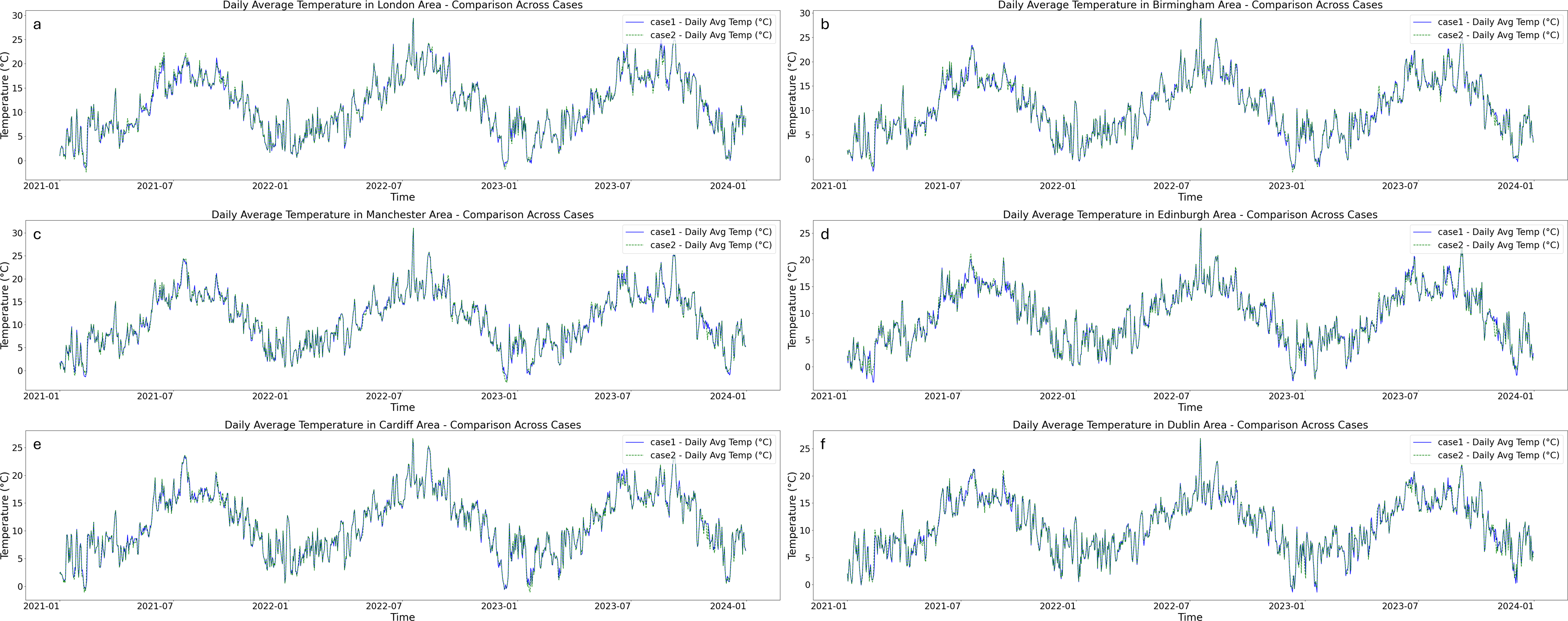}
\caption{The daily average temperature variations along three years under case 1 and case 2 over (a) London, (b) Birmingham, (c) Manchester, (d) Edinburgh, (e) Cardiff and (f) Dublin.}
\label{fig:C2}
\end{figure}

\begin{figure}[htb]
\includegraphics[width=12cm]{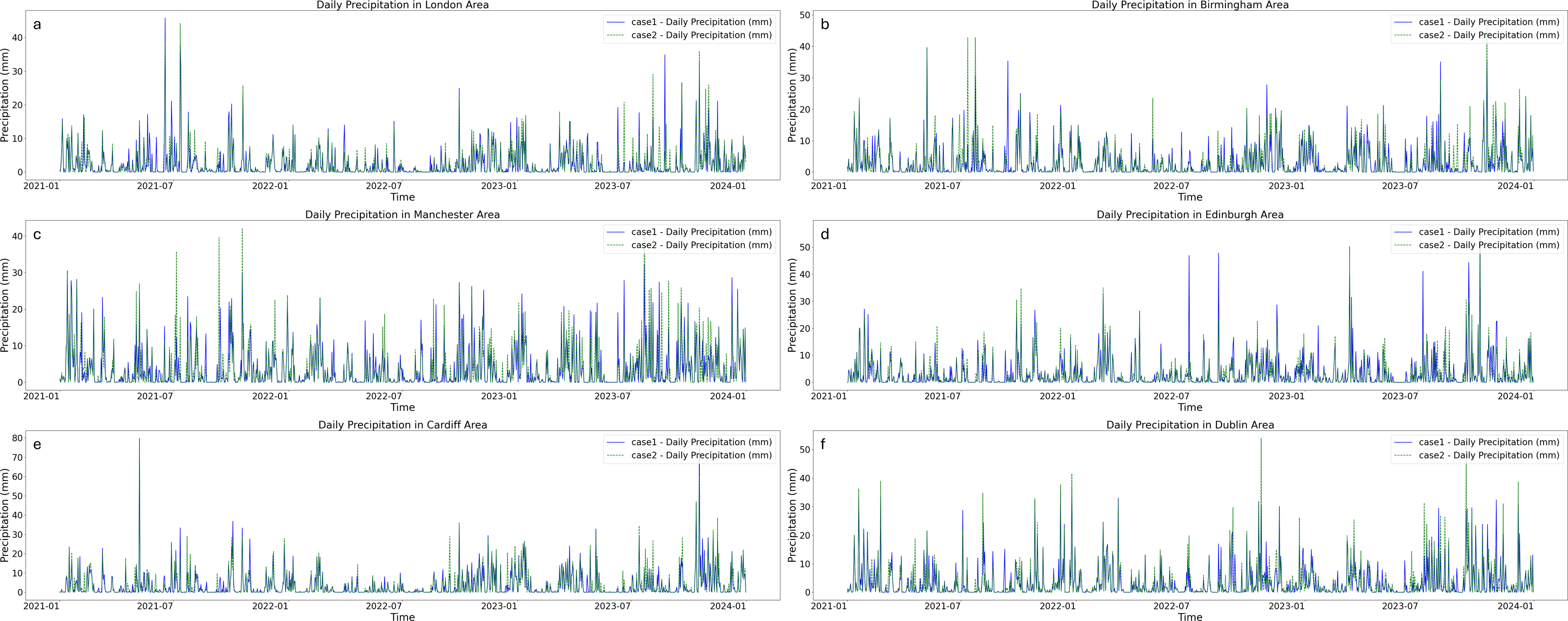}
\caption{The daily average precipitation variations along three years under case 1 and case 2 over (a) London, (b) Birmingham, (c) Manchester, (d) Edinburgh, (e) Cardiff and (f) Dublin.}
\label{fig:C3}
\end{figure}

\begin{figure}[htb]
\includegraphics[width=12cm]{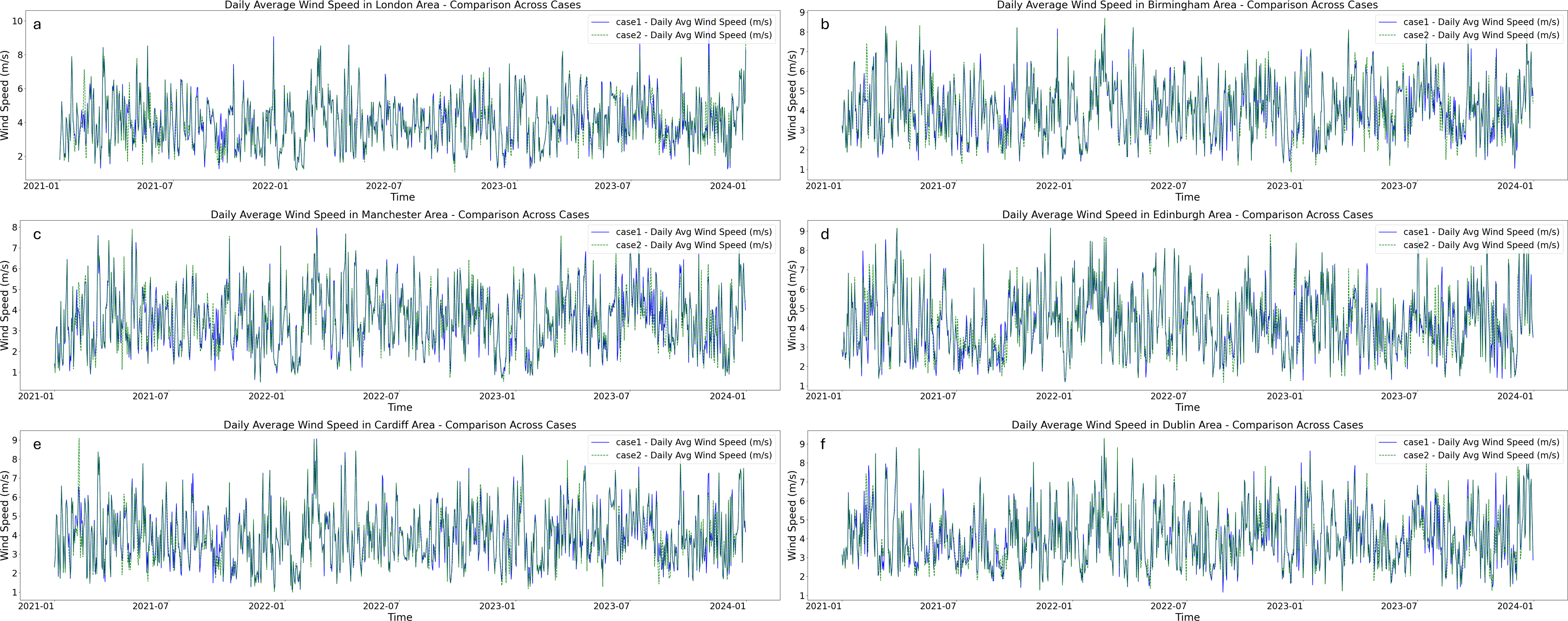}
\caption{The daily average wind speed variations along three years under case 1 and case 2 over (a) London, (b) Birmingham, (c) Manchester, (d) Edinburgh, (e) Cardiff and (f) Dublin.}
\label{fig:C4}
\end{figure}

\begin{figure}[htb]
\includegraphics[width=12cm]{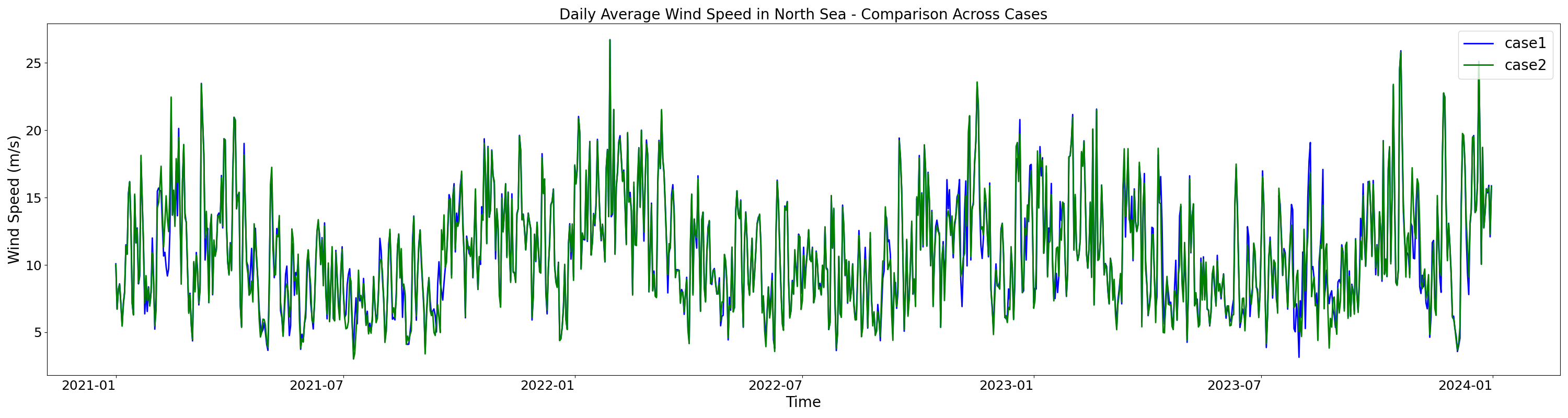}
\caption{The 150 m daily average wind speed variations along three years under case 1 and case 2 over the North Sea.}
\label{fig:C5}
\end{figure}

In this part, more comparisons between Case 1 and Case 2 are provided in terms of the average difference of daily minimum and daily maximum temperatures (Fig. \ref{fig:C1}), the daily average temperature (Fig. \ref{fig:C2}), precipitation (Fig. \ref{fig:C3}) and wind speed (Fig. \ref{fig:C4}) variations over six cities as well as the 150 m daily average wind speed variations over the North Sea (Fig. \ref{fig:C5}).

\backmatter

\section*{Declarations}

\bmhead{Acknowledgments}
This work has received funding from the UK Engineering and Physical Sciences Research Council (grant number: EP/Y016297/1). We express great appreciation to the National Center for Atmospheric Research for providing WRF model, Copernicus Climate Change Service for providing the EAR5 data, Group for High Resolution Sea Surface Temperature for providing the GHRSST data, the National Meteorological Information Center for providing weather station measurement data, the National Renewable Energy Laboratory (NREL) for providing power curves of reference wind turbines, and the NASA EOSDIS Land Processes Distributed Active Archive Center (LP DAAC), U.S. Geological Survey (USGS)/Earth Resources Observation and Science (EROS) Center for providing MODIS MCD12Q1.061 land cover type data. We extend our heartfelt gratitude to Ting Zhang, Bo Tian, Dhritiraj Sengupta, Lei Zhang and Yali Si for providing global offshore wind turbine dataset 2021 and Simon Waldman, Munro Peter and Forster Rodney for providing the plausible 2050 offshore wind locations.

\bmhead{Data availability}
The ERA5 data are available at the European Centre for Medium-Range Weather Forecasts, \url{https://www.ecmwf.int/en/forecasts/dataset/ecmwf-reanalysis-v5}. The GHRSST V4.1 data are available at the Physical Oceanography Distributed Active Archive Center (PO.DAAC), \url{https://podaac.jpl.nasa.gov/dataset/MUR-JPL-L4-GLOB-v4.1}. The weather station measurement data are available at the National Meteorological Information Center \url{https://data.cma.cn/ai/#/detail?id=5}. The MODIS MCD12Q1.061 data are available at U.S. Geological Survey (USGS) \url{https://lpdaac.usgs.gov/products/mcd12q1v061/}. The global offshore wind turbine dataset 2021 is available at \url{https://figshare.com/articles/dataset/Global_offshore_wind_farm_dataset/13280252}. The plausible 2050 offshore wind locations are available at \url{https://zenodo.org/records/10259046}. The power curves of reference wind turbines are available at \url{https://nrel.github.io/turbine-models/index.html}.

\bmhead{Code availability}
The WRF V4.3.3 model is available at \url{https://github.com/wrf-model/WRF/releases/tag/v4.3.3}. Due to the high volume of the simulation results, the raw data are unavailable to be shared, while the namelist used to produce the experiments will be available at \url{https://github.com/warwick-icse/WRF_Climate}. 

\bmhead{Author contributions}
R.L. proposed and refined the concept, designed and implemented the whole framework including the WRF simulation, performance verification, designed and implemented the dataset analysis and visualization, and wrote the paper draft. J.Z. contributed to conceptualization, interpreting the results, and refining the manuscript. X.Z. contributed to acquiring funding, analysing and  interpreting the results, and refining and editing the manuscript. X.Z. supervised the overall study and investigation.




\end{appendices}


\bibliography{bibliography}


\end{document}